\documentclass[aps,twocolumn,showpacs]{revtex4} 
\usepackage{graphicx}
\usepackage{amsmath}
\usepackage{amssymb}

\def\cv{c_{\rm v}} 
\def\rs{r_{\rm s}}
\def\GeV{\ {\rm GeV}} 
\def\cm{{\rm cm}}

\def\rtilde{\tilde{r}}
\def\rhos{\rho_{\rm s}}

\def\Msun{M_\odot}
\def\Mmw{M_{\rm MW}}
\def\rhosm{\rho_{\rm sm}}
\def\rhosun{\rho_\odot}
\def\rhosub{\rho_{\rm sub}}

\def\Prho{P(\rho)}

\begin{document}

\title{Galactic Substructure and Direct Detection of Dark Matter}

\author{Marc Kamionkowski}
\affiliation{California Institute of Technology, Mail Code
130-33, Pasadena, CA 91125; kamion@tapir.caltech.edu}
\author{Savvas M. Koushiappas} 
\affiliation{Los Alamos National Laboratory, Mail Stop B227, Los
  Alamos, NM 87545; smkoush@lanl.gov }


\begin{abstract}
We study the effects of substructure in the Galactic halo on
direct detection of dark matter, on searches for
energetic neutrinos from WIMP annihilation in the Sun and
Earth, and on the enhancement in the WIMP annihilation rate in the halo. 
Our central result is a probability distribution
function (PDF) $P(\rho)$ for the local dark-matter density.  This
distribution must be taken into account when using null
dark-matter searches to constrain the properties of dark-matter
candidates.  We take two approaches to calculating the PDF.  
The first is an analytic model that capitalizes on the
scale-invariant nature of the structure--formation hierarchy 
in order to address early stages in the hierarchy (very small 
scales; high densities). 
Our second approach uses simulation-inspired
results to describe the PDF that arises from lower-density
larger-scale substructures which formed in more recent stages
in the merger hierarchy.  The distributions are skew positive,
and they peak at densities lower than the mean density.  The
local dark-matter density may be as small as 1/10th the
canonical value of $\simeq0.4\,\GeV~\cm^{-3}$, but it is probably 
no less than $0.2\,\GeV~\cm^{-3}$.
\end{abstract}

\maketitle

\section{Introduction}

The nature of dark matter remains a mystery.  Among the plethora
of dark-matter candidates, there are two classes that are
sufficiently promising to motivate major experimental searches.
The first is a weakly interacting massive particles (WIMP)
\cite{Jungman:1995df,BHS05},
which may arise in supersymmetric extensions of the
standard model or in theories that include universal extra
dimensions (UED) \cite{Hooper:2007qk}.  The other leading candidate is the
axion \cite{axionreviews}, hypothesized to solve the strong-CP problem.

WIMPs may be detected indirectly through observation of gamma
rays or cosmic-ray positrons, antiprotons, and/or antideuterons
produced when WIMPs annihilate in the Galactic halo or in the
halos of some extragalactic systems.  WIMPs might also be detected
via observation of energetic neutrinos produced by annihilation
of WIMPs that have accumulated in the Sun and/or Earth.
There is also an effort to detect dark-matter particles in 
low-background experiments via detection of the $O(100\,{\rm keV})$
energy they impart to nuclei from which they elastically
scatter.  Likewise, dark-matter axions are being sought directly
by conversion to photons in resonant-cavity experiments
\cite{axionexperiments}.  If
dark matter is composed of WIMPs, we may also get some clue
to its nature from forthcoming LHC accelerator experiments
\cite{lhc}.

Predictions for the event rates for any of these detection
schemes depend on the distribution of dark matter in the
Galactic halo.  The flux of gamma and cosmic rays from
WIMP annihilation depends on an integral of the square of the 
dark matter density over volume, 
while the rates for energetic neutrinos and direct detection 
depend only on the local dark-matter density.  In early 
calculations, it was assumed simply that dark matter was
smoothly distributed in the Galactic halo, with either an
isothermal or NFW \cite{NFW96} radial density profile, and with a local
dark-matter density fixed by Milky Way dynamics to be
$\rhosun\simeq0.4~\GeV~\cm^{-3}$.

It has become clear in recent years, from theory and
N-body simulations, that the distribution of dark matter in the
Galactic halo is not likely to be perfectly smooth, and that
some of the dark matter in the Galactic halo will be distributed
into subhalos with a variety of sizes \cite{substructure}.  
This substructure hierarchy may extend to very small scales
\cite{Boehm,Green:2003un,Diemand:2005vz,Diemand:2007qr,Ando:2005xg}.
For WIMPs in supersymmetric and UED models, the cutoff scale of
the power spectrum is in the range $[10^{-6} -
10^{2}\, ]M_\oplus$ \cite{Chen:2001jz,Profumo:2006bv}.
For axions, the growth of perturbations is suppressed on scales
smaller than $(m_a H)^{-1/2}$ \cite{Hu:2000ke,Boyle:2001du}, the
geometric mean
of the inverse of the axion mass and the Hubble constant; for example,
for an axion of mass
$m_a\sim10^{-5}$~eV, the cutoff scale corresponds to about
$10^{-11}\,M_\oplus$.

The implications of such substructure for searches for cosmic
rays from WIMP annihilation in the halo have been explored
extensively in the literature \cite{subgamma}.  
Simply stated, if a dark matter halo contains substructure, then the volume
integral of the density squared is increased by some boost 
factor $B$.  For example, Ref.~\cite{Giocoli:2007gf} recently claimed
$B\sim10^3$ enhancements in the direction of the Galactic
anticenter, although Ref.~\cite{Berezinsky:2005py} finds an
enhancement $B\simeq2-5$. An analytic approach presented in 
Ref.~\cite{Strigari:2006rd} also finds smaller values of $B$, of 
order $B \sim 10$. 

The implications of substructure for {\it direct} detection of dark
matter have, however, been comparatively neglected.  This is a
serious omission, as substructure implies fluctuations in the local
dark-matter density, and these fluctuations imply an uncertainty
in the predicted rates for direct-detection experiments and for
energetic-neutrino searches.  A simple but illustrative
example places all of the dark matter in subhalos of density $B$
times the mean density.  In this case, the total annihilation
rate is enhanced by a boost factor $B$, but the
probability that the Solar System lives in such a subhalo is only
$B^{-1}$.  If $B\gg 1$, this probability is small, implying dire
prospects for direct-detection, an alarming conclusion that
warrants further investigation.  

Few past studies focused on the implications of
substructure for the local dark-matter density. 
Some focused on implications derived from numerical simulations 
\cite{Moore:2001vq,Helmi:2002ss}, thus limited by resolution effects.  
Others looked at the contribution of tidal streams to the local density 
\cite{Stiff:2001dq,Freese:2003tt}. 

The goal of this paper is to begin addressing questions such as: 
What are the possible values of the local density?  How small
can they be?  What are the most probable values? 
To answer these questions, we calculate a probability
distribution function (PDF) $P(\rho)$ for the local dark-matter
density. This PDF accounts for fluctuations in the local
density due to substructure. The mean of this distribution
is $\simeq0.4~\GeV~\cm^{-3}$, the canonical value determined
from dynamics, but we find that the local dark-matter density may 
be as small as 1/10th, but probably no less than half this canonical value. 
The PDF we calculate will have 
to be convolved with the results of null searches to infer
constraints to particle-dark-matter parameters.
In addition, the PDF can be used to assess what a substructure enhancement 
in the halo WIMP annihilation rate implies for the local dark matter density, 
and vice-versa. 

Our first approach, in Section II, uses an {\it ansatz} about
the survival fraction for subhalos that form early at high
densities.  General
arguments will be used to bracket the plausible range of PDFs,
and also the plausible lower limit to the local dark-matter
density.  This analytic approach capitalizes on the
scale-invariant nature of the hierarchical formation of galaxies
to extrapolate simulation results to the highest-density,
smallest-scale substructures that form earliest and that are
beyond the reach of simulations.  We then perform a second
calculation that assumes substructures with NFW profiles
are distributed in the Milky Way halo.  This second approach
describes the local PDF due to larger-scale substructures which
have formed from more recent stages in the merger hierarchy, and
it thus complements the first approach.  In Section III, we
improve upon this latter calculation by using ingredients on
subhalo mass functions and concentration parameters taken
from numerical simulations.  We compare our work to past
literature in Section IV, and we state our conclusions in Section V.

\section{Analytic Model} 

In hierarchical structure formation, small halos
collapse first, and they then merge to form more massive
structures.  Because each halo virializes to roughly 200 times the
mean cosmological density at collapse, the halo density
generally increases with decreasing subhalo mass.  

However, a dark matter halo is unlikely to remain completely intact as
it merges into larger structures in the hierarchy, and a
significant fraction of the mass of each subhalo will be stripped
as it is embedded into larger halos.  As a result, a significant 
fraction of the mass of any halo in the hierarchy will be
smoothly distributed, rather than contained in substructure.  
Nevertheless, there will be
some fraction that is contained in subhalos of higher density, originating from
an earlier stage in the hierarchy, and some fraction of those will be entrapped  
into even smaller and denser sub-subhalos, etc., all the way
down to the smallest scale in the hierarchy.  It is important to
note that 
N-body simulations are unlikely to ever be able to resolve the
complete hierarchy of substructures, from $10^{12}$~$M_\odot$ to
$10^{-10}\, M_\odot$, or even smaller, and so some analytic
approach is required.  Likewise, extended-Press-Schechter calculations of 
substructure keep track of the {\it most} massive subhalo that
each mass element occupies, not the {\it least} massive subhalo
in the hierarchy, which is the step in the hierarchy that
determines the local density. 

\subsection{Local density probability distribution function}

We describe structure formation as a series of
steps in a hierarchy, where the first stages consist of the
lowest-mass and densest subhalos, which then merge in subsequent
steps in the hierarchy to form more massive structures of lower
density.  Note that this model simplifies by assuming that each
halo has a uniform density.  However, as we show below, 
the qualitative results we obtain are essentially
unaltered, at least at high density, if we assume more
realistic density profiles, such as isothermal or NFW profiles.

Let $f(\rho_1)$ be the fraction of 
mass, per logarithmic density interval centered on
$\rho_1$, that is not yet locked up in halos of higher density. 
Let $F(\rho_1)$ then be the fraction of
the mass today in the Milky Way halo that exists at density
greater than $\rho_1$.  Then $F(\rho_1)$ is related to $f(\rho_1)$
through
\begin{equation}
     \frac{dF}{d\rho_1} = -\frac{f}{\rho_1} (1-F).
\label{eqn:Feqn}
\end{equation}
In other words, $-(dF/d\rho_1)d\rho_1$ is the fraction of the mass
in the halo today that has a density in the interval $\rho_1\rightarrow
\rho_1 + d\rho_1$, and this is $f/\rho_1$ times $1-F$, the fraction of
mass not yet locked up in higher-density subhalos.

Strictly speaking though, $\rho_1$ is the density that a halo
{\it would} have if the mass was uniformly distributed
throughout the halo.  The
{\it true} density $\rho$ of the smoothly distributed
component of any subhalo is reduced because some of the
mass is in subhalos of higher density.  A given halo has a total
mass $M$ and occupies a volume $V=M/\rho_1$, but the mass in
subhalos is $MF(\rho_1)$, and these subhalos occupy a volume
\begin{equation}
     V_{\rm sub} = \int_{\rho_1}^{\rho_{\rm max}} \, d\rho_1'\,(-dF/d\rho_1')
     [M/\rho(\rho_1')].
\end{equation}
The smoothly distributed matter has mass $M[1-F(\rho_1)]$, and it
occupies a volume 
$V - V_{\rm sub}$, 
and so the true density of
the smoothly distributed component of a halo of mean density
$\rho_1$ is
\begin{equation}
     \rho(\rho_1) = \rho_1 \frac{ 1-F(\rho_1)}{ 1
     +\rho_1\int_{\rho_1}^{\rho_{\rm max}} \, \frac{dF}{d\rho_1'} \,
     d\rho_1' \frac{1}{\rho(\rho_1')}}.
\label{eqn:integralequation}
\end{equation}
Thus, for example, locally in the Milky Way, the density
$\rhosun\simeq0.4\, \GeV~\cm^{-3}$ is the value determined from
dynamics, but $\rho$, the smoothly-distributed component, may
be smaller, reduced because some of the mass is in subhalos. 

Eq.~(\ref{eqn:integralequation}) is an integral equation for
$\rho(\rho_1)$.  To
solve it, we differentiate the expression for $\rho(\rho_1)$
with respect to $\rho_1$ to obtain a differential equation,
\begin{equation}
     \frac{d\rho}{d\rho_1} = \frac{\rho^2}{ \rho_1^2
     [1-F(\rho_1)]}.
\label{eqn:diffeq}
\end{equation}
This can then be integrated to obtain
\begin{equation}
     \rho(\rho_1) = \left\{ \int_{\rho_1}^{\rho_{\rm max}}
     \frac{d\rho_1'}{ (\rho_1')^2 [1-F(\rho_1)]} \right\}^{-1}.
\label{eqn:integralexpression}
\end{equation}
This result could have also been obtained simply by discretizing
and summing the contributions of each density interval to the volume.

To recapitulate, we postulate a survival fraction $f(\rho_1)$ for
halos of mean density $\rho_1$.  We obtain the fraction of the
Milky Way mass smoothly distributed in halos of mean density
$\rho_1$ by solving Eq.~(\ref{eqn:Feqn}).  We then obtain the true
density $\rho$ of the smoothly distributed component of halos
of density $\rho_1$ from Eq.~(\ref{eqn:integralexpression}).  The
fraction of mass in the Milky Way with density in the interval
$\rho\rightarrow d\rho$ is then
$(-dF/d\rho)=(-dF/d\rho_1)(d\rho/d\rho_1)^{-1}$.  The final step
is then to simply note that the density probability distribution
function $P(\rho)$ that we seek is actually the fraction of the
Milky Way {\it volume}, rather than mass, at density $\rho$, and
this is
\begin{eqnarray}
     P(\rho) &=& \frac{1}{V} \frac{dV}{d\rho} \nonumber \\
     &=& \frac{
     (1/\rho) (-dF/d\rho)}{ \int_{\rho_\odot}^{\rho_{\rm
     max}} d\rho_1' [1/\rho(\rho_1')] (-dF/d\rho_1')}.
\label{eqn:pdf}
\end{eqnarray}

With this result, we have mapped the problem of determining
$P(\rho)$ to the problem of determining the survival fraction $f(\rho_1)$.
In principle, $f(\rho)$ can be determined with
simulations, but the appropriate simulations have not yet
been performed.  Thus, here we will make some educated guesses,
based partially on estimates from existing simulations, and then explore the
implications of these guesses for $P(\rho)$.  

N-body simulations
of cosmological formation of Milky-Way-like halos suggest that
only about [10--20]\% of the Milky-Way-halo mass is in its 10
largest progenitors \cite{andrew}. This fraction could potentially 
be larger as halos in future numerical simulations get more highly resolved. 
For example, the survival fraction of early forming subhalos (thus tightly 
bound) is higher ($\sim 40 \%$ of their mass survives \cite{Diemand:2007qr}) 
than recently 
forming subhalos; therefore if subhalos are resolved at higher 
densities, the fraction of mass in substructure as measured in simulations 
could increase. 
If we assume [10--40]\% of the Milky Way halo mass is in subhalos 
with a density $\sim 1/10$th that of the mean density of the Milky Way, 
then this implies very roughly that $f(\rho_1)\simeq 0.04-0.16$, at least for the
most recent stage in the hierarchy ($\rho_1\simeq\rhosun$).
Therefore, for the remainder of this calculation we will 
consider a range $f(\rhosun)=0.05-0.2$. 

How then does $f(\rho_1)$ vary for higher masses?  Before
answering, we first note that the density and mass of the inner 10
kpc of the Milky Way, where we live, suggest a formation
redshift $z\sim10$ for the inner 10 kpc of the Milky Way halo.
At these and higher redshifts, the Universe is Einstein-de
Sitter.  Moreover, the effective spectral index for primordial
perturbations spans the range $-2.6 \lesssim n_{\rm eff}
\lesssim -2.1$ for mass scales $10^{11}\, M_\odot \gtrsim M
\gtrsim 10^{-6}\,M_\oplus$, corresponding to the range of
substructure mass scales we are considering here.  If we
approximate $n_{\rm eff}\simeq$constant, then in the appropriate
range of stages in the hierarchy, structure formation is scale
invariant.  One reasonable guess is thus that
$f(\rho_1)=$constant, that the hierarchy of substructures in the
Galactic halo is scale invariant.

However, what is more likely, for several reasons, is that
$f(\rho_1)$ decreases with increasing $\rho_1$.  First of all,
subhalos in the earlier stages of the hierarchy undergo more
orbits in their parent halos between the time they form and
today, and this increases the efficiency with which they will be
stripped \cite{Berezinsky:2005py}.  Secondly, smaller subhalos
may be disrupted by tidal interactions with stars \cite{hsz}.
Third, the effective spectral index $n_{\rm
eff}\rightarrow -3$ for earlier stages in the hierarchy.  Thus,
earlier in the hierarchy there is less time for a subhalo to
virialize fully before it becomes enveloped by the next larger
halo in the hierarchy.  To model these effects, we thus consider
$f(\rho_1) =f(\rhosun)(\rho_1/\rhosun)^{-\alpha}$ with $\alpha>0$.

The differential equation in Eq.~(\ref{eqn:Feqn}) is readily
solved for these $f(\rho_1)$ parameterizations.  For $f(\rho_1)=$constant,
$F(\rho) = 1 -(\rho_1/\rho_{\rm max})^f$.  For example, if we take
$\rho_{\rm max}/\rhosun=1000$, then for $f(\rho_1)=0.05$, 0.1, or
0.2, we find that in the Milky Way, $F(\rhosun)=0.29$, 0.5,
and 0.75, respectively; the remaining $(1-F)$ mass is in the smooth
component.  Thus, if $f=0.1$, then half of the 
mass of the Milky Way halo at radii $r\sim8$~kpc is smoothly
distributed, and the other half is in subhalos.  A larger $f$
reduces the smooth component, as does a larger $\rho_{\rm
max}$.  As we show below, a power-law $f(\rho_1)$ increases
the smooth component.  The density of the smooth component
can be estimated to be $(1-F)\rhosun$.  More
precisely, solution of Eq.~(\ref{eqn:diffeq}) gives the density
of the smooth component as $\rho(\rhosun)=0.74$, 0.55, and
0.3, respectively.

For a power-law $f(\rho_1)=f(\rhosun)(\rho_1/\rhosun)^{-\alpha}$
(and $\alpha>0$),
\begin{equation}
     F(\rho_1) = 1 -\exp \left\{ -\frac{f}{\alpha} \left[ 1 -
     \left( \frac{\rho_1}{\rho_{\rm max}} \right) \right]
     \right\},
\end{equation}
and if $\alpha$ is not too small ($\alpha \ge 0.5$ ), 
the dependence on the cutoff
density $\rho_{\rm max}$ disappears.  In that case,
$F(\rhosun)\simeq 1-\exp[-f(\rhosun)/\alpha] \simeq
f(\rhosun)/\alpha$, where the last approximation valid for
$f(\rhosun) \ll \alpha$.  One guess for $f(\rho_1)$ is that it is
inversely proportional to the formation time $t\propto
\rho^{-1/2}$ for halos of density $\rho$.  If so, then
$F(\rhosun)\simeq 2 f(\rhosun)$ is the fraction of the Milky Way
mass that is in subhalos, roughly 20\% for $f(\rhosun)=0.1$.  A
stronger dependence on $\rho_1$ would result in a smaller
$F(\rhosun)$.  It is then straightforward to calculate the PDF
$P(\rho)$ to find power-law dependence $P(\rho) \propto
\rho^{-(2+\alpha)}$ for densities above the smooth density.

\begin{figure}[t]
\resizebox{!}{8.5cm}{\includegraphics{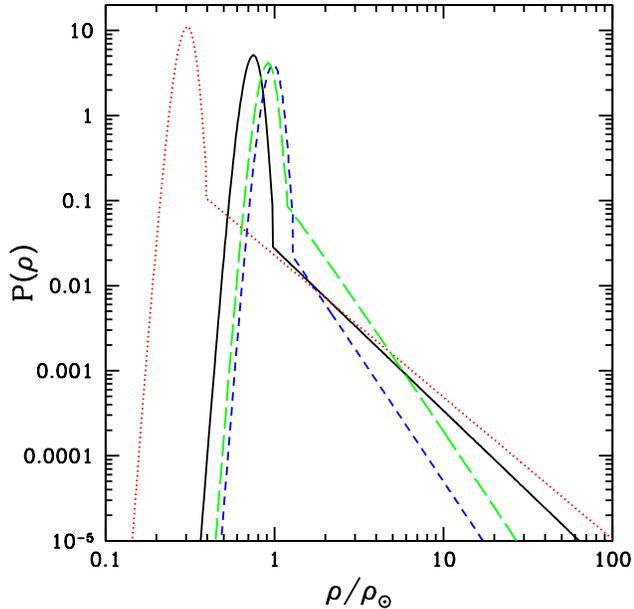}}
\caption {\small The local dark-matter-density probability
     distribution function $P(\rho)$ for the analytic model, as
     a function of the density $\rho$
     scaled by the mean density $\rhosun$,
     for $\{f(\rhosun),\alpha\}=\{0.05,0\}$ (black solid curve),
     $\{0.2,0\}$, (dotted red curve), $\{0.05,1\}$, (short-dash
     blue curve) $\{0.2,1\}$ (long-dash green curve).  We smooth
     the Dirac delta function for the smooth component to a
     Gaussian of rms one-tenth the smooth-component density.
     The power-law tails are due to subhalos.
}
\label{fig:logPDF}
\end{figure}

\subsection{Results and Discussion}

To illustrate, Fig.~\ref{fig:logPDF} shows the PDFs, on a
log-log plot, for four combinations of the parameters
$f(\rhosun)$ and $\alpha$ in the model.  The distributions feature a
high-density tail (close to a power-law), 
due to the fraction $(1-F)$ of the
mass that is in substructure.  Also shown
is a smooth component at a density $\rho_{\rm sm} < \rhosun$;
for purposes of illustration, we smooth the Dirac delta-function
dependence of this smooth component to a Gaussian of rms
one-tenth the smooth-component density.  The Figure shows that
as $f(\rhosun)$ is increased, the amplitude of the high-density
tail is increased at the expense of the smooth component.  We
also see that increasing $\alpha$ increases the smooth-component
density while decreasing the amplitude of the high-density tail.

Table~\ref{table:table} provides numerical results for the
smooth-component density, the fraction of the volume occupied by
the smooth component, and the annihilation enhancement $B$
(discussed in more detail below) for PDFs $P(\rho)$ parameterized by the
survival fraction $f(\rhosun)$ and the survival-fraction
power-law index $\alpha$.

Here are some comments and general conclusions from these results so far:

(1) If $f(\rhosun)$ is roughly 
$f\lesssim 0.2$ as suggested by numerical simulations, then the
reduction in the smooth-component density is no less than 30\%
the mean density.  The implied fractional uncertainty in the local
dark-matter density is thus not too much larger than that
(roughly factor of two) implied by
uncertainties in the stellar/gas contribution to the local
rotation curve, or the uncertainties that arise if we allow for
a flattened halo (which generally increase the local dark-matter
density).

\begin{table}[htb]\footnotesize
\begin{center}
\begin{tabular}{c|c|c|c|c}
$f(\rhosun)$ & $\alpha$ & $\rho_{\rm smooth}$ & smooth fraction &
$B$ \\
\hline
0.05 & 0 & 0.75 & 95\% & 47 \\
\hline
0.1 & 0 & 0.56 & 91\% & 88 \\
\hline
0.2 & 0 & 0.3 & 83\% & 156 \\
\hline
0.05 & 0.5 & 0.95 & 97\% & 3.9 \\
\hline
0.1 & 0.5 & 0.89 & 94\% & 6.8 \\
\hline
0.2 & 0.5 & 0.78 & 88\% & 12 \\
\hline
0.05 & 1 & 0.98 & 98\% & 1.3 \\
\hline
0.1 & 1 & 0.96 & 95\% & 1.6 \\
\hline
0.2 & 1 & 0.91 & 91\% & 2.1 \\
\hline
\end{tabular}
\caption{The density $\rho_{\rm smooth}$ of the smooth
component, the fraction of the volume occupied by the smooth
component, and the annihilation enhancement $B$ for density
distributions $P(\rho)$ parameterized by the survival-fraction
amplitude $f(\rhosun)$ and the survival-fraction power-law index
$\alpha$.}
\label{table:table}
\end{center}
\end{table}

(2) The fraction of the volume occupied by the smooth component
is larger than the fraction of the mass in the smooth component,
as the higher-density components occupy correspondingly less
volume.  As a result, in most of the models, the density in the
vast majority of the halo volume is the density of the smooth
component.

(3) We have assumed that each halo and subhalo has a
uniform density. More realistically, the density of each halo
and subhalo will decrease with radius, perhaps with an NFW
profile, which has a density that depends on radius $r$ as
$\rho\propto r^{-1}$ in the inner regions.  The volume in
such a halo changes with density as $(dV/d\rho)\propto
\rho^{-3}$ for $\rho\rightarrow\infty$.  This falls with density
more rapidly at large $\rho$ than $P(\rho)\propto
\rho^{2+\alpha}$ as long as $\alpha<1$.  Thus, if $\alpha<1$,
the high-density scaling of our $P(\rho)$ will be unaltered by
convolving our halo density distribution with NFW profiles.

By contrast, this analytic model is weakest perhaps at low
densities (larger--scale substructures), where details of the
low--density outer regions of
merging substructures may be important in determining the local
density, the density of what we have called the smooth
component.  To address this shortfall, we continue in the next
subsection to calculate the halo PDF in a different model, and one
that will then preview the numerical results that we present later.

\subsection{Discrete subhalo populations}

In this Section we consider another illustrative model, one in
which earlier generations of subhalos are subsumed and then
smoothly distributed in radius into their larger host halos.  In this
case, the Milky Way halo is seen as an amalgamation of smaller
halos, distributed in mass with some mass function $dn/dM$
(e.g., a Press-Schechter-like mass function, or one taken from
simulations).  Furthermore, each of these subhalos has an NFW
density profile.  In this scenario, there is no smooth component of 
dark matter distributed in the Milky Way halo; instead, the whole halo 
is made up of adjacent individual subhalos. Although there may
be a wide variety of halo
masses in this scenario, they all have comparable mean
densities, and mean densities comparable to the Milky Way
density.  If this is the case, then the subhalos fill the entire
volume of the Milky Way.  Such a scenario neglects the
halos-in-halos problem, but it may more accurately describe the
PDF due to larger-scale substructures, which arise from the most
recent stages in the merger hierarchy.

Consider now the volume of radius $R$ occupied by one of these
subhalos.  If the Milky Way mass were uniformly distributed, the
matter in this volume would have some uniform density
$\bar\rho$. However, the subhalo has its own NFW density profile,
\begin{equation}
     \rho(r) = \frac{\rho_s}{(r/r_s)(1+r/r_s)^2},
\end{equation}
where $r_s$ is the scale
radius.  This scale radius is related to the concentration
parameter $\cv$ through $R=\cv r_s$; i.e., halos with larger
$\cv$ are more centrally concentrated.  The characteristic
density $\rho_s$ is related to $\bar\rho$, $R$, and $\cv$
through $\bar\rho = 3 \rho_s f(\cv)/\cv^3$, where $f(\cv)=\ln \cv -
\cv/(1+\cv)$.

\begin{figure}[t]
\resizebox{!}{8.5cm}{\includegraphics{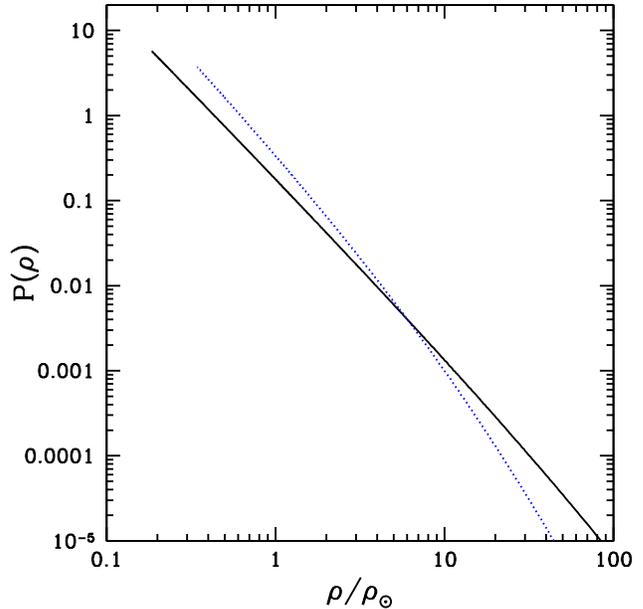}}
\caption {\small The local dark-matter-density probability
     distribution function $P(\rho)$, for the discrete-subhalo
     model, as a function of the density $\rho$
     scaled by the mean density $\rhosun$, 
     for $\cv=10$ (black solid curve) and $\cv=2$ (dotted red curve).
}
\label{fig:lognfw}
\end{figure}

The fraction of the volume at density $\rho$ in this model is
$P(\rho)=(1/V)(dV/d\rho) = (3/R^3)r^2 (dr/d\rho)$.  It can be calculated
analytically, but the expressions are algebraically unwieldy and
unilluminating.  The important thing is that $(dV/d\rho)\propto
\rho^{-3}$ for $\rho \gg \rho_s$ and $(dV/d\rho) \propto
\rho^{-2}$ for $\rho \ll \rho_s$.  The results for $P(\rho)$ in
this model are shown in Fig.~\ref{fig:lognfw}.  These models
predict a broad range of $\rho$, but the median densities are
$\rho_{\rm med}=0.35\,\rhosun$ and $0.58\,\rhosun$ for $\cv=10$
and 2, respectively.  The 95\% C.L. lower limits to $\rho$ are
$\rho_{95} = 0.20\, \rhosun$ and $0.36\,\rhosun$ for $\cv=10$
and 2, respectively.

Note that nowhere in this discussion did we specify the subhalo
mass or mass function.  The results apply as long as all of
the subhalos have the same concentration parameter and virial
density, independent of their mass.  More realistically, there
will be a range of concentration parameters and virial
densities, and this will be explored in the next Section.  Note
also that we have assumed no smooth component, but more
realistically the outer parts of each subhalo will be tidally
stripped to provide a smooth component.  This can be seen
roughly by noting that the density of the NFW profiles at the
maximum radius is roughly 0.19--0.34 the mean density.  This
will also be explored in the next Section.

\subsection{Annihilation enhancement due to substructure}

As we mention in the Introduction, substructure in the halo
implies an enhancement to the dark-matter annihilation rate. 
For example, a $10^{-4} M_\oplus$ halo has a characteristic
density which is roughly 1000 times the local dark-matter
density.  If all the dark matter in the halo today resided in
such subhalos, then the annihilation flux would be boosted by a factor
of 1000.  However, the probability for the Solar System to be
in such a subhalo would then be only 0.1\%.

The analytic models developed here allows us to evaluate the
enhancement of the annihilation rate due to substructure, and relate
it to the local dark-matter density. The enhancement in the annihilation rate, 
over the smooth dark-matter distribution is
\begin{eqnarray}
     B &=&\frac{\int\,\rho^2 dV}{\int\, \rhosun^2\, dV}
     \nonumber \\
     & =& \int\, P(\rho) \frac{\rho^2}{\rhosun^2} d\rho.
\label{enhancement}
\end{eqnarray}

If the survival fraction $f(\rho_1)$ has a power-law dependence
$f(\rho_1)\propto \rho_1^{-\alpha}$, then there will be a
contribution $\sim f(\rhosun)(\rho_{\rm
max}/\rhosun)^{1-\alpha}/(1-\alpha)$ to the enhancement from the
high-density tail in $P(\rho)$.  Thus, a large annihilation
enhancement requires $\alpha<1$, $f(\rhosun)$ not too small ($f(\rhosun) \ge 0.2)$, 
and
$\rho_{\rm max} \gg \rho$.  For example, we estimate that for
$f(\rho)=0.2$ (a constant) and $\rho_{\rm max} \simeq
1000\,\rhosun$, the enhancement factor will be $B \sim200$.  The
enhancement factor will be roughly proportional to $f(\rhosun)$,
and it will decrease, possibly sharply, as $\alpha$ is increased
from zero.  Our numerical results find that $B$ decreases to
values $B\lesssim10$ for $\alpha=0.5$ and $f(\rhosun)\lesssim
0.2$.  It is also important to note that if the power-law
exponent of $\alpha$ is such as to allow a large enhancement $B$,
then the value of that enhancement is likely to depend strongly
on the cutoff density $\rho_{\rm max}$.

For the case where the Milky Way halo is composed of discrete subhalos with 
no smooth component, the annihilation rate in this halo is enhanced over that in the
uniform-density halo by a factor
\begin{eqnarray}
     B(\cv) &=& \frac{4 \pi \int\, r^2\, dr\, [\rho(r)]^2}{ 4 \pi
     \int\, r^2\, dr\, \bar\rho^2} \nonumber \\
     &=& \frac{1}{3} \frac{\cv^3 g(\cv)}{[f(\cv)]^2},
\end{eqnarray}
where $g(\cv) = (1/3) [1 -(1+\cv)^{-3}]$.  This enhancement
factor is $B(\cv=1)\simeq 2.6$, it grows to $B(\cv=10)\simeq
50$, and grows roughly as $\cv^3/9[\ln \cv]^2$ for $\cv \gg 1$, 
in broad agreement with the results presented in
Ref.~\cite{Strigari:2006rd}.

In summary, we conclude that: 

(1) Very large (i.e.,
$\gg10$) enhancements to the annihilation rate require a
survival fraction $f$ that is roughly constant with density.  In
other words, the survival fraction for a halo must be largely
{\it in}dependent of its formation time.  If earlier halos are
less likely to survive, then a large annihilation enhancement
requires a survival fraction today that is large, and perhaps
too large to be consistent with numerical simulations of halo
formation.  This is consistent with Ref.~\cite{Berezinsky:2005py}, who
claim that survival fractions of the earliest halos are
$\sim0.1\%-0.5\%$ and an annihilation fraction  $B\simeq2-5$.

(2) The annihilation enhancement generally increases
at the expense of the local density. For example, in the
$\alpha=0$ model, an annihilation enhancement $B \approx 150$ implies a
reduction by a factor of three in the local dark-matter
density.  The annihilation enhancement will generally increase
and the local smooth component will also generally
decrease for larger $\rho_{\rm max}$. 

(3) Strictly speaking, the PDF derived here is for the dark-matter density
{\it locally} (or at some other specified point), and the
annihilation enhancement $B$ is the
enhancement in the annihilation rate in a local volume.  It is
expected that the PDF, as well as $B$, will vary with radius in
the Galactic halo.  Since the central higher-density regions of
the halo presumably formed earlier, the amount ot substructure should be
reduced there (i.e., a larger smooth fraction and smaller $B$),
and conversely for larger radii (smaller smooth fraction and
larger $B$).  These trends are consistent with the enhancement
factors found, for example, in Ref.~\cite{Giocoli:2007gf}, which
calculate the annihilation intensities as a function of
observation direction. 

\section{Simulation-inspired results}

We now proceed to build on the analytic calculation in the
previous Section by implementing a substructure fraction, a subhalo
mass function, and a range of concentration parameters taken
from simulations.

We assume that when averaged over time, the solar neighborhood
has a mean density  $\rhosun \approx 0.4 \GeV \, \cm^{-3}$.  In
a halo that contains substructure, this density can be split into a smooth
component, and a component that arises from the presence of
subhalos,
\begin{eqnarray}
\label{eq:rhosun}
\rhosun &=&\rhosm + \rhosub \nonumber \\
              &=& \rhosm + \int_{\epsilon}^{\eta} M \frac{dn}{d \ln M} d \ln M .
\end{eqnarray}
Here, $dn/d \ln M$ is the number density of objects per
logarithmic mass interval, and the integral is performed over 
a mass range, $\epsilon \le M \le \eta \le \Mmw$.  Strictly
speaking, the mass function depends on the primordial power
spectrum and upon the physics of halo merging and stripping of
halos.  

Numerical simulations find that the mass
function can be approximated over a wide mass range by a
power law, $dn/d \ln M \sim M^{-\beta}$, with $\beta \sim 0.9$
\cite{Diemand:2007qr,2005MNRAS.359.1537R,2004MNRAS.352..535D},
with indications that a similar form remains down to sub-solar
subhalos \cite{Diemand:2006ey,Berezinsky:2005py}.  We
normalize the mass function so that a fraction $\xi$ of the
local mean dark-matter density is in objects with masses between
$\epsilon \approx 10^{-10}\Msun$ and $\eta \approx 10^{12} \Msun$.
This implies that the value of the smooth component is $\rhosm =
( 1 - \xi ) \rhosun$.  The upper limit $\eta$ of the integration
must always be less than the mass of the Milky Way. However, the
lower integration limit $\epsilon$  depends on the physics of the
dark-matter particle and on small-scale structure assemblage.

We model each subhalo with a two-parameter NFW profile, taking
the two parameters to be the mass and the concentration
parameter, and using a virial radius (maximum radius) defined by assuming 
the density of each halo is 200 times the matter density at the 
redshift of formation. 
Each subhalo is then assigned a concentration that is selected
from a $P(\cv)$ log-normal distribution about a mean determined by the 
value of $\sigma(M)$ and the evolution of linear perturbations, 
and with a scatter
$\sigma[\log (\cv)] =0.14$ inferred from numerical simulations
\cite{Bullock:1999he,Wechsler:2001cs}. In this model, the mean 
concentration is a weak function of mass, $\cv \sim M^{-\gamma}$, where 
$\gamma \approx 0.13$ for scales near $M_\star$ at $z\approx 0$, and 
scales as $\cv \approx 33 ( M / 10^8 M_\odot)^{-0.06}$ for halos with 
masses $M \le 10^{8} M_\odot$. 

We calculate the density probability distribution function as
\begin{equation}
\label{eq:prho}
     P(\rho) =  \int_{\epsilon}^{ \eta} 
     \frac{dn}{d \ln M} \, \frac{dv(M,\rho)}{d \rho} \,d \ln M .
\end{equation}
Here, $dv(M,\rho)/d\rho = \int [dv(M(\cv))/d \rho ]\, P(\cv) \, d \cv$, where 
$v(M,\rho) = 4 \pi (\rs \rtilde_c)^3 / 3$ is the
volume in a halo of mass $M$, where $\rtilde_c$ is obtained by solving 
\begin{equation}
\label{eq:rc}
     \rtilde_c ( 1 + \rtilde_c)^2 = \rhos / \rho. 
\end{equation}
In solving Eq.~(\ref{eq:rc}), we assume that halos can only
have a finite size, given by $\rtilde_{\rm max} = \cv$. If the
solution is such that $\rtilde < \cv$, then the volume with
density greater than $\rho$ is simply $4 \pi \rs^3 \rtilde^3 /
3$, while if $\rtilde > \cv$, then the volume is given by $4 \pi
\rs^3 \cv^3 / 3$.  However, in order to approximate the effects
of tidal interactions, Eqs.~(\ref{eq:prho}) and (\ref{eq:rc}) 
are always solved for $\rho > \rhosm$; i.e., we assume the
tidal radius of each subhalo is defined as the radius where the
density of the halo is equal to the local smooth density
component.

\begin{figure}[t]
\resizebox{!}{8.5cm}{\includegraphics{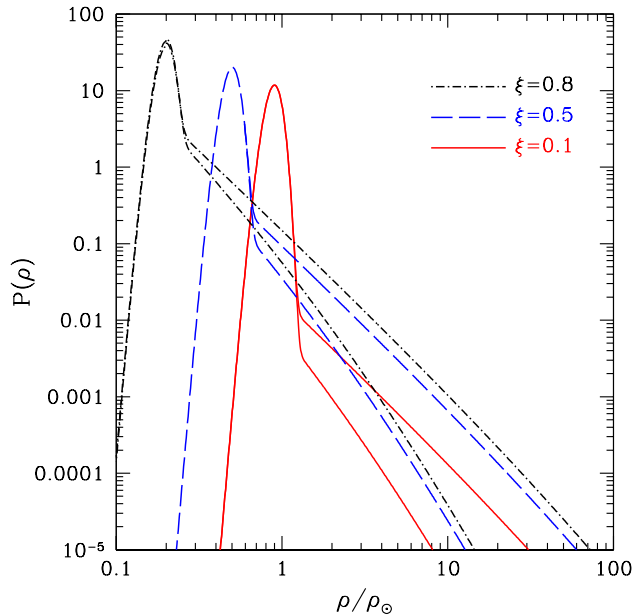}}
\caption {\small The probability distribution function in the
     solar neighborhood for the simulation-inspired
     calculation. The solid, long-dash and dot-dash curves 
     correspond to $\xi=0.1$, 0.5, and 0.8 respectively. As in
     Fig.~\protect\ref{fig:logPDF}, we smooth 
     the Dirac delta function for the density value of the smooth component 
     with a Gaussian of rms a tenth of the smooth-component density.
     The upper and lower curves for each value of $\xi$ show the 
     range of contribution of the subhalo population that arises from 
     uncertainties in the subhalo mass function, as well as the subhalo 
     population (see text).}
\label{fig:fig1}
\end{figure}

Fig.~\ref{fig:fig1} shows the density probability distribution function as 
a function of $\rho$, expressed as a fraction of $\rhosun$.
Different line types depict the PDFs that corresponds to
$\xi=$0.1, 0.5, and 0.8.  As before, we smooth the Dirac delta
function at the value of the smooth density
by a Gaussian with an RMS one-tenth the smooth component density. 
The range that corresponds to the subhalo population for each $\xi$ 
represents the spread in $\Prho$ that
arises from a spread in the lower integration limit $\epsilon$
and in the power-law exponent $\beta$ of the subhalo mass
function.  More precisely, for each value of $\xi$, the upper
curve corresponds to a subhalo power-law index $\beta=1.1$ and
integration limits $\epsilon = 10^{-10}~\Msun$ and
$\eta=10^{-9}~\Msun$; i.e. a steep subhalo mass function with most
of the population in objects of extremely small mass.  The lower
curve is obtained by flattening the subhalo mass function to a
power-law exponent $\beta = 0.7$ and integration limits
$\epsilon = 10^{10}~\Msun$ (since we know the LMC exists) and
$\eta=10^{12}~\Msun$, roughly speaking the Milky Way mass; i.e.,
this is a model with only very massive substructure. 

The value of $\xi$ fixes the value of $\rhosm$, and
densities $\rho > \rhosm$ come from subhalos. The slope of $\Prho$ is set in
Eq.~(\ref{eq:prho}) by the slope of $v(\rho)$.  At densities
where $\rtilde \ll 1$, the volume within which the density is
greater than a particular value is $ v \sim \rtilde^3 \sim
\rho^{-3}$. For lower densities, probing the outer regions of a
halo---i.e., $\rtilde \gg 1$---the volume scales as $v \sim
\rtilde^3 \sim \rho^{-1}$.  Thus, we expect a slope for $\Prho$
of $-2$ for low densities, and a slope of $-4$ as the density
increases.

The sharp transition between the smooth component
and the subhalos in Fig.~\ref{fig:fig1} results from the
assumption of a pure power law for the dark-matter profile in subhalos
\cite{NFW96}. In reality, the transition will be much more gradual,
reflecting the fact that subhalos are embedded in the potential
well of their host Milky Way halo.  As $\xi \rightarrow 1$, the
smooth component approaches zero. There is still a
lower limit, $\rho \gtrsim 0.04 \GeV \cm^{-3}$, attained at a
value $\xi \gtrsim 0.92$, to the local halo density that arises
from the overlap of the outskirts of subhalos.  These
calculations thus imply that there is significant probability
for the local halo density to be below $\rhosun$, but also that
there is likely a minimum possible value to the local density.
Simulations (e.g., Ref.~\cite{Diemand:2005vz}) suggest a value
$\xi\simeq0.5$, but further investigation is required to
determine this important parameter more precisely and robustly.

The annihilation enhancement in this approach can be estimate as in 
Ref.~\cite{Strigari:2006rd}. Namely, in a halo of mass $M$ and 
with a cut-off in the subhalo mass function at $m_0$, the 
boost factor is the solution of an integral equation that takes 
into account the halo-in-halo problem, and is approximated 
by $B \approx 0.1 [ (M/m_0)^{0.13} - 1 ]$. For a cut-off in the power 
spectrum at microhalo scales, $m_0 \approx 10^{-6} M_\odot$, 
and a Milky Way mass of $M\approx 10^{12} M_\odot$, $B\approx 20$. 
The weak dependence of the
boost factor to the cut-off scale of the subhalo mass function is an 
outcome of the flatness of the dark-matter power spectrum, which manifests 
itself in the concentration--mass relationship which enters the
boost-factor calculation (see Ref.~\cite{Strigari:2006rd}).

\section{Discussion}

There is a vast literature on galactic substructure,
much of it discussing the implications for dark-matter
detection.  Most of this work focuses on the annihilation boost
factor, but there are few papers that discuss the
implications for direct detection.  The primary focus of these
papers is then on the effects of substructure on the local
dark-matter velocity distribution, with less attention to the
possible implications for the local dark-matter density.  Here
we review some of the work that discusses the implications of
substructure on the local dark-matter density.

In Ref.~\cite{Helmi:2002ss}, the authors use 
numerical simulations to investigate the
effects of substructure on the local velocity distribution
and on the local density.  The conclusion of that paper is that
most of the local dark matter is smoothly distributed
(illustrated in their Fig.~5, which resolves structure down to
$\sim10^7\,\Msun$).  The fraction of mass that is bound in
halos less massive than $10^7\, \Msun$ is less than 4\%; this is
the fraction of the mass that could have survived the tidal
field of the Milky Way.  They thus conclude that it is highly
unlikely that a good fraction of the halo is in Earth-mass
subhalos.  However, their calculation does not consider the
halos-in-halos problem: i.e, their mass function $dN/dm$ keeps
track only of the most massive halo in which a given particle
resides, not the {\it least} massive subhalo.  Put another way,
their calculation counts only the Earth-mass halos that did {\it
not} get incorporated into more massive halos, and it disregards
those that did, assuming simply that they were completely
disrupted.  Still, this paper does argue that substructures at
later stages in the merger hierarchy get largely erased. 
Although isolated Earth-mass substructures may survive in a
smooth Milky Way halo (as pointed out also in
Ref.~\cite{Moore:2001vq}), when scaled to earlier generations in the hierarchy, 
the conclusions of this paper 
imply that
Earth-mass subhalos are likely to be disrupted when they merge
into subsequent stages in the hierarchy (e.g., into
$1000\,M_\oplus$ halos).  If so, then $f(\rho)$ will be very
small ($ f(\rho) \ll 0.2)$).

A local dark matter density PDF $P(\rho)$ is calculated in
Ref.~\cite{Stiff:2001dq},
but it is a different distribution.  In
particular, their calculation assumes that the vast majority of
the local dark matter is smoothly distributed, and that only
$1\%-5\%$ of the local dark matter may be in substructure.  The density
$\rho$ that appears in their PDF is thus the density of this
additional substructure component, which they assume comes from
the latest subhalos accreted onto the Milky Way halo.  Their
mass function does extend to small masses, but they do not
consider halos-in-halos.  Ref.~\cite{Freese:2003tt} is similar in
spirit, but considers in particular the effects of the tails of
the Sagittarius dwarf.

The approach that most closely resembles ours is perhaps that in
Ref.~\cite{Moore:2001vq}.  They recognize the scale-invariant nature of
the problem---that is, that there may be halos in halos---but
then note that the resolution limits of their simulation
prohibits them from making definitive claims about earth-mass
objects.  Their calculations indicate that the singular cores of
subhalos may always survive, even if most of the mass from
a given subhalo is stripped.  They provide as an example a
subhalo orbiting at 20 kpc in the Milky Way in which only 0.3\%
of the initial mass remains after four orbits, but then show
that the survival fraction may be as high as 40\% for an orbit
at 40 kpc.  The simulations of Ref.~\cite{Moore:2001vq} make the
important point that tidal tails broaden rapidly, and this
justifies the assumption of our first analytic model that
matter stripped from a subhalo is rapidly smoothly distributed
in the new larger halo.

This assumption is also justified by the recent results presented 
in Ref.~\cite{Vogelsberger:2007ny}. In this work, the authors proposed a 
technique for calculating the fine-grained phase space structure in 
dark matter halos from cosmological N-body simulations. 
In demonstrating the effectiveness of this 
new method, they studied the evolution of the dark matter density 
that arises from streams in NFW-like potentials, and found that 
a very large number of streams ($\sim 10^5$) may potentially 
be present in the solar neighborhood. If each stream has a diameter of 
order $\sim$kpc, it means that their entanglement and evolution has an 
effect in the local {\it smooth} dark matter density, and thus do not address 
the potential enhancement on much smaller scales.

\section{Conclusions and Challenges}

It is evident from simulations and analytic arguments that
some fraction of the local Milky Way dark matter may be in subhalos.
The implications of substructure for indirect detection of WIMPs
have been studied broadly, with the conclusion that there may be
large enhancements in annihilation rates over the rates
predicted assuming a smooth halo.  The implications for {\it
direct} searches have, however, been largely overlooked.  This
is a possibly serious omission, as one consequence of substructure is that
the local density will be smaller than the smoothly-distributed
local density usually assumed.

We have taken a few first steps to understand the
implications of substructure for the local density.  Our central
goal is a calculation of the PDF $P(\rho)$ for the local
dark-matter density $\rho$.  This $P(\rho)$ will need to be
taken into account when interpreting the implications of null
dark-matter searches for constraints to the particle-dark-matter
parameter space (e.g., couplings and/or elastic-scattering cross
sections).  

We considered two simple scenarios for substructure:  In
the first, early generations of very dense subhalos survive with
some probability.  The advantage of this approach is that if
subhalo survival fractions can be measured in simulations for
recently merged subhalos, the results might be extrapolated to
the much earlier generations (much smaller and denser subhalos)
that may be below the resolution of simulations.  

In the second approach, the halo is assumed to consist of
recently formed subhalos, each with an NFW profile.  This
approach provides a simple, albeit approximate, way to
understand the effects of larger-scale substructure, from more
recent stages in the merger hierarchy, on the PDF.  
We then pursued this approach further using
subhalo mass functions and concentration-parameter distributions
taken from simulations.  This approach does not take into
account the possible contribution of subhalos within
subhalos.  In principle, a complete solution for the PDF,
including substructures on the largest and smallest scales, can
be obtained by convolving our two calculations.  This, however,
will be left for future research.

Substructure scenarios that yield larger
annihilation enhancements generally imply a smaller local
dark-matter density.  Very large annihilation enhancements require that
the very densest substructures, which generally form earlier,
must survive through all later generations of structure
formation.  If earlier substructures are less likely to survive
than more recent substructures, then a very large annihilation
enhancement is unlikely.  

So, how small can the local density be?  The smallest local
density in the models we surveyed was one tenth the canonical
value of $0.4\, \GeV\,\cm^{-3}$, usually assumed for a smoothly
distributed halo.  This small value was obtained from our
simulation-inspired result using what we believe to be an overly
conservative estimate of the smooth fraction, which we obtained
by truncating NFW subhalos when their density falls below
the mean halo density.  More realistic values for the smooth
fraction are probably in the range of 50\%-80\%, which would
then correspond in our simulation-inspired
results to the $\xi=0.8$ and $\xi=0.5$ distributions,
respectively, shown in Fig.~\ref{fig:fig1}.  The smallest local
density in our analytic model for early substructure was 0.3,
obtained using the high value, $f(\rhosun)=0.2$, and assuming a
constant $f(\rho_1)$.  If, however, $f(\rhosun)$ is lower, then
the local density is increased.  More importantly, it is quite
likely that $f(\rho)$ decreases with $\rho$, and if so, the
local density is increased to $\sim0.8$ times its canonical
value, even for $f(\rhosun)=0.2$, and even higher for smaller
$f(\rhosun)$.  A combination of our two approaches, to take into
account both low- and high-density substructures, will likely
show that the local density is no less than half the canonical
value.

The volume of the halo probed during a three-year
direct-detection experiment is very small, and so the halo
density $\rho$ that we have been discussing can be safely
assumed to be the density averaged over the duration of a
direct-detection experiment. 
However, the rate for production of
energetic neutrinos from WIMP annihilation in the Sun or Earth
lags behind the rate for capture of WIMPs from the halo by an
equilibration time $t_{\rm eq}$
\cite{Griest:1986yu,Kamionkowski:1991nj} that can vary
considerably with the WIMP's mass and elastic-scattering and
annihilation cross sections; typical values might be $t_{\rm
eq}^{\odot}\sim 5\times 10^7$ yr and $t_{\rm eq}^\oplus\sim10^{10}$ yr
\cite{Kamionkowski:1994dp}.  An energetic-neutrino search thus
probes the halo density averaged over a much larger volume (for the 
latest bounds on the flux of high-energy neutrinos from annihilations 
in the Earth and the Sun see \cite{Desai:2004pq,Ackermann:2005fr,
Achterberg:2006jf}).The halo density averaged over this volume will 
thus have a PDF that will be narrower than that
for direct detection, and the minimum density will
be closer to the mean halo density. However, even though initially it was 
assumed that the velocity distribution function of dark matter 
particles mirrors that in free space (thus justifying the use of 
a Maxwell-Boltzmann distribution) \cite{Gould:1991}, it is possible that the 
velocity distribution function is surpressed in the low-velocity 
tail due to the effects of solar capture and WIMP diffusion in the 
solar system due to the presence of other planets \cite{Gould:1999je,Lundberg:2004dn}. 
This surpression manifests itself as a reduction in the number 
density of dark matter particles near the Earth for WIMPs with 
masses greater than few hundred GeV. Nevertheless, there is a
possibility that for low-mass WIMPs, given the longer equilibration 
time for the Earth relative to that for the Sun, the annihilation 
signal from the Earth could be boosted relative to that for the Sun, 
if the Solar System passed through a very dense subhalo at a time
$t_{\rm eq}^\odot <t<t_{\rm eq}^\oplus$ ago.  We leave a more
detailed calculation of the density smoothed
over timescales relevant for energetic-neutrino searches to
future work \cite{KK}.

The PDF $P(\rho)$ allows us to evaluate also the boost factor in
the annihilation rate.  It must be emphasized, though, that the
boost factor $B$ we have considered is the boost only in the {\it
local} annihilation rate (per unit volume).  Strictly, speaking,
the PDF may vary from one position in the halo to another.
Thus, for example, if we assume that the halo is spherically
symmetric (after averaging over substructure fluctuations), then
the PDF will be a function $P(\rho;r)$ of Galactocentric radius
$r$, as well as the density $\rho$.  The mean density of this
distribution, as a function of
$r$, will be the density of the
smooth NFW profile that best fits the rotation curve.
Qualitatively, we expect that the high-density tail will be more
pronounced (more substructure) at larger $r$ and less pronounced
(less substructure) at smaller $r$.  The PDF $P(\rho;r)$ can
then be used to determine the boost factor as a function of
position in the halo; again, we expect $B(r)$ to vary with $r$
and to generally increase with $r$.  This $B(r)$ will be
essential to compute annihilation fluxes along various
lines of sight through the Galaxy.  We encourage future authors
to describe the effects of substructure in terms of $P(\rho;r)$,
or at least in terms of $B(r)$, as this may facilitate
comparison between the conclusions of different studies.

\begin{acknowledgments}
We acknowledge useful conversations with A.~Benson, S.~Habib, K. Heitmann, 
G.~Jungman, D.~Nagai, and D.~Reed.  SMK thanks Caltech for
hospitality while this work was being conceived.  Work at LANL
was carried out under the auspices of the NNSA of the
U.S. Department of Energy at Los Alamos National Laboratory
under Contract No. DE-AV52-06NA25396.   This work was supported
at Caltech by DoE DE-FG03-92-ER40701, NASA NNG05GF69G, and the
Gordon and Betty Moore Foundation.
\end{acknowledgments}

\end{document}